\documentclass{article}
\usepackage{spconf,amsmath,graphicx}
\usepackage{amssymb,amsfonts}
\usepackage{algorithm}
\usepackage{algpseudocode}
\usepackage{breqn}
\usepackage{mathtools}
\usepackage{tikz}
\usetikzlibrary{shapes,arrows}
\usepackage{subcaption}
\usepackage{cite}
\usepackage{xcolor,soul}
\usepackage{hyperref}
\usepackage{siunitx}
\input{mysymbol.sty}
\newtheorem{problem}{\hspace{0pt}\bf Problem}

\renewcommand{\baselinestretch}{0.97} 

\DeclareSIUnit{\symbol}{\text{symb}}

\title{Joint channel estimation and data detection in massive MIMO systems based on diffusion models}
\name{Nicolas Zilberstein$^{\star}$, Ananthram Swami$^{\dagger}$, Santiago Segarra$^{\star}$\thanks{Research was sponsored by the Army Research Office and was accomplished under Cooperative Agreement Number W911NF-19-2-0269. The
views and conclusions contained in this document are those of the authors
and should not be interpreted as representing the official policies, either ex-
pressed or implied, of the Army Research Office or the U.S. Government.
The U.S. Government is authorized to reproduce and distribute reprints for
Government purposes notwithstanding any copyright notation herein. Emails: \{nzilberstein, segarra\}@rice.edu., ananthram.swami.civ@army.mil.}}
\address{$^{\star}$Rice University, USA \hspace{1.5cm}$^{\dagger}$DEVCOM Army Research Laboratory, USA}
         
\begin{document}
\ninept
\maketitle
\begin{abstract}

We propose a joint channel estimation and data detection algorithm for massive multilple-input multiple-output systems based on diffusion models.
Our proposed method solves the blind inverse problem by sampling from the joint posterior distribution of the symbols and channels and computing an approximate maximum a posteriori estimation.
To achieve this, we construct a diffusion process that models the joint distribution of the channels and symbols given noisy observations, and then run the reverse process to generate the samples.
A unique contribution of the algorithm is to include the discrete prior distribution of the symbols and a learned prior for the channels.
Indeed, this is key as it allows a more efficient exploration of the joint search space and, therefore, enhances the sampling process.
Through numerical experiments, we demonstrate that our method yields a lower normalized mean squared error than competing approaches and reduces the pilot overhead.
\end{abstract}
\begin{keywords}
Joint channel estimation and data detection, score-based generative models, Langevin diffusion, blind inverse problems.
\end{keywords}
%

%%%%%%%%%%%%%%%%%%%%%%%%%%%%%%%%%%%%%
%%%Introduction%%%%%%%%%%%%%%%%%%%%%%
%%%%%%%%%%%%%%%%%%%%%%%%%%%%%%%%%%%%%
% \vspace{-0.1in}
\section{Introduction}\label{S:intro}
% Massive multiple-input multiple-output (MIMO) systems are fundamental in the development of modern and forthcoming wireless communications~\cite{mimoreview1}.
% An essential attribute underpinning augmented data rates and spectral efficiency resides in the large number of antennas that base stations are equipped with, which allows concurrent support of several users.
% Although this is fundamental for moving from the fifth to the sixth generation of cellular communications~\cite{6g}, it also entails many challenges such as \emph{designing low-complexity MIMO detection algorithms} and \emph{the development of accurate channel estimation methodologies}.
% These two problems at the physical layer are our focus in this paper.
Massive multiple-input multiple-output (MIMO) systems are pivotal for advancing wireless communication~\cite{mimoreview1}. 
These systems feature a large number of antennas at base stations, enabling simultaneous support for multiple users. 
While crucial for the transition to 6G cellular networks~\cite{6g}, MIMO systems pose challenges, such as achieving low-complexity detection and precise channel estimation. 
This paper focuses on these physical layer issues.

% Exact MIMO detection is an NP-hard problem~\cite{Pia2017MixedintegerQP}.
% Given $N_u$ users and modulation of $K$ symbols, the exact maximum likelihood (ML) estimator has an exponential decoding complexity $\mathcal{O}(K^{N_u})$.
% Thus, the combination of a large number of users and higher-order modulation schemes makes the ML estimate computationally intractable even for moderately-sized systems. 
% Furthermore, it relies on perfect channel state information (CSI), which in general is not available.
% Hence, the design of a channel estimation scheme that recovers accurate CSI using a reduced pilot overhead is mandatory for a successful deployment of the MIMO detector.
% However, this is challenging in massive MIMO due to the high dimensionality of the system and the constraint on the number of pilot signals, and thus, it is an open research problem~\cite{balevi2020high}.

Exact MIMO detection is NP-hard~\cite{Pia2017MixedintegerQP}. 
With $N_u$ users and  $\kappa$-symbol modulation, the maximum likelihood estimator's decoding complexity is exponential $\mathcal{O}(\kappa^{N_u})$, rendering it infeasible for sizable systems. Additionally, it relies on perfect channel state information (CSI), usually unavailable. 
Hence, an efficient channel estimation method is crucial, but it is challenging in massive MIMO due to the high dimensionality of the system and the constraint on the number of pilot signals~\cite{balevi2020high}.
As a consequence, a joint optimization approach (still NP-hard) to estimate both the data and the channel is necessary to tackle these issues in high-dimensional systems~\cite{vikalo2006efficient, yi2020deep}.
Different approximate solutions based on gradient descent have been proposed for single-input multiple-output systems~\cite{castaneda2017vlsi}.
Recently, new methods have been proposed for massive MIMO.
In~\cite{song2021joint}, the authors proposed an algorithm that iteratively solves a relaxed version of the maximum a posteriori joint channel estimation and data detection (MAP-JED) problem that exploits the sparsity of the channel, while in~\cite{song2021soft} a solution based on deep neural networks, and more specifically, algorithmic unfolding, was proposed.
In this work, we introduce a novel approach based on \emph{diffusion models}, also known as score-based generative modeling.

Diffusion models, known for their versatility, have excelled across diverse applications~\cite{song2021scorebased}. 
These models transform training data into white Gaussian noise (the forward process) and then run the reverse transformation (the generative process) to generate new data. 
Leveraging their ability to generate high-dimensional data from unknown probability distributions, diffusion models have achieved state-of-the-art results in various areas, such as denoising and inpainting in image processing~\cite{kawar2022denoising,chung2023diffusion}. 
In wireless technology, they have been used to derive a massive MIMO detector based on Langevin diffusion~\cite{zilberstein2022annealed, zilberstein2023}, and for channel estimation~\cite{arvinte2022}, both of which have shown remarkable results.
However, these methods are tailored for problems with known forward operators and may not apply to \textit{blind} inverse problems, where the operator is unknown. 
Recent advances in image processing aim to address these challenges~\cite{levac2022accelerated, Chung_2023_CVPR}, to provide more comprehensive solutions.

In this work, inspired by the above results, we propose a framework to solve the joint massive MIMO channel estimation and data detection problem based on score-based generative modeling.
To achieve this, we construct a diffusion process that models the joint distribution of the channels and symbols given noisy observations and then sample it by running the reverse diffusion.
Our proposed method allows us to incorporate prior information on each variable to enhance the sampling process.

\vspace{0.2mm}
\noindent
{\bf Contribution.}
The contributions of this paper are twofold:\\
1) We propose an algorithm for solving the joint massive MIMO detection and channel estimation problem based on diffusion models, which allows us to incorporate both the discrete prior of the symbols and a learned prior for the channel.\\
2) Through numerical experiments, we analyze the behavior of our method for different hyperparameter settings and demonstrate the performance gain of our joint solution compared to separate channel estimation and data detection.
%
%
%%%%%%%%%%%%%%%%%%%%%%%%%%%%%%%%%%%%%%%%%%%%%%%%%%%%%%%%%%%
%%%System model and problem formulation%%%%%%%%%%%%%%%%%%%%%%
%%%%%%%%%%%%%%%%%%%%%%%%%%%%%%%%%%%%%%%%%%%%%%%%%%%%%%%%%%%
\vspace{-0.1in}
\section{System model and problem formulation}
\vspace{-0.1in}
%
% \blue{SS: Is Section 2 an exact copy of your previous paper? If yes, please try to change it a bit.}\nz{ready}
We focus on the uplink of a massive MIMO communication system.
We consider $N_u$ single-antenna transmitters or UEs that transmit pilots and data to a receiving base station equipped with $N_r$ antennas, under a block-fading scenario with a coherence time of $K = P + D$ time slots; $P$
time slots are reserved for pilots and $D$ time slots are used for payload data.
Under this scenario, the forward model for the MIMO system is defined as
\begin{equation}\label{E:mimo_model}
	\bbY = \bbH \bbX + \bbZ,
\end{equation}
where $\bbH \in \mathbb{C}^{N_r \times N_u}$ is the channel matrix, each $\{[\bbZ]_{:j}\}_{j=1}^K \sim \mathcal{CN}(\bb0, \sigma_0^2\bbI_{N_r})$ is a vector of complex circular Gaussian noise, $\bbX = [\bbX_P, \bbX_D] \in \mathcal{X}^{N_u \times K}$ is the vector of transmitted data, where $\bbX_P\in \mathcal{X}^{N_u \times P}$ corresponds to the pilots while $\bbX_D \in \mathcal{X}^{N_u \times D}$ is the data, and $\mathcal{X}$ is a finite set of constellation points, and $\bbY \in \mathbb{C}^{N_r \times K}$ is the received vector.
We consider quadrature amplitude modulation (QAM) throughout this work with symbols normalized to attain unit average power. 
All the users transmit with the same modulation and each symbol has the same probability of being chosen by each of the  $N_{u}$ users. 
Moreover, we assume that $\bbH$ is unknown while $\sigma_0^2$ is known at the receiver.
Under this configuration, the joint channel estimation and data detection problem can be {stated} as follows.
\vspace{-0.03in}
\begin{problem}\label{P:main} \emph{
    Given observations $\bbY$ and pilots $\bbX_{P}$ following~\eqref{E:mimo_model}, find the MAP estimate of both $\bbX_D$ and $\bbH$.}
\end{problem}
\vspace{-0.03in}
Given that $\bbZ$ in~\eqref{E:mimo_model} is a random variable, a natural way of solving Problem~\ref{P:main} is to search for $\{\bbH, \bbX_D\}$ that maximizes its \emph{posterior} probability given the noisy observations $\bbY$ and pilots $\bbX_P$.
Hence, the Bayes' optimal decision rule can be written as
\begin{align}\label{eq:map}
	\{\hat{\bbH}_{\mathrm{MAP}}, \hat{\bbX}_{D_{\mathrm{MAP}}}\} &= \argmax_{\substack{\bbX_D \in \mathcal{X}^{N_u\times D}, \\ \bbH \in \mbC^{N_r\times N_u}}}\,\, p(\bbX_D, \bbH|\bbY,\bbX_P)
	\\ 
    \nonumber&= \argmax_{\substack{\bbX_D \in \mathcal{X}^{N_u\times D}, \\ \bbH \in \mbC^{N_r\times N_u}}}\,\, p_{\bbZ}(\bbY - \bbH\bbX)p(\bbX)p(\bbH).
\end{align}
% \vspace{-4mm}
%
The problem in~\eqref{eq:map} is known as MAP-JED.
Since we assume that the symbols' prior distribution is uniform among the constellation elements and the measurement noise $\bbZ$ is Gaussian, the MAP-JED formulation boils down to the following optimization problem
\begin{equation}\label{eq:ml}
	\{\hat{\bbH}_{\mathrm{MAP}}, \hat{\bbX}_{D_{\mathrm{MAP}}}\} = \argmin_{\substack{\bbX_D \in \mathcal{X}^{N_u \times D}, \\ \bbH \in \mbC^{N_r\times N_u}}}\,\, ||\bbY - \bbH\bbX||^2_2 - \log p(\bbH),
\end{equation}
% \vspace{-4mm}
%
\noindent This problem is NP-hard due to the discrete nature of the finite constellation constraint $\bbX_D \in \mathcal{X}^{N_u}$ and the non-convexity of the product between decision variables.
Thus, several schemes have been proposed in the last decades to provide efficient approximate solutions to Problem~\ref{P:main}, as mentioned in Section~\ref{S:intro}.
In this paper, we propose to solve Problem~\ref{P:main} by (approximately) sampling from the posterior distribution in~\eqref{eq:map} using an annealed Langevin dynamic.

%%%%%%%%%%%%%%%%%%%%%%%%%%%%%%%%%%%%%
%%%Our approach%%%%%%%%%%%%%%%%%%%%%%
%%%%%%%%%%%%%%%%%%%%%%%%%%%%%%%%%%%%%
\vspace{-0.1in}
\section{Joint channel estimation and data detection based on Langevin dynamics}
\vspace{-0.07in}

In Section~\ref{subsec:langevindyn}, we briefly introduce the Langevin diffusion and score-based generative modeling.
Then, in Section~\ref{subsec:posterior}, we describe our algorithm for joint posterior sampling based on Langevin dynamics, and detail the expression of the score functions involved in the diffusion process.
\vspace{-0.1in}
\subsection{Langevin diffusion and posterior sampling}
\label{subsec:langevindyn}

In general, a diffusion process is a continuous-time Markov process on the variable $\bbX \in \reals^d$ that solves the Ito equation 
\begin{align}\label{eq:diffusion}
    \text{d}\bbX_t &= \bbf(\bbX_t, t)\text{d}t + \bbg(t)\text{d}\bbW_t,
\end{align}
\noindent where $\bbW$ is a standard $d$-dimensional Brownian motion, $\bbf(\bbX_t,t)$ and $\bbg(t)$ are the drift and diffusion term respectively, which are assumed to be Lipschitz
continuous for all t.
The Langevin diffusion is a particular type of diffusion process, which is obtained when $f(\bbX_t, t) = \nabla_{\bbX_t}\log p(\bbX_t)$ and $g(t) = \sqrt{2 \tau}$~\cite{pavliotis_book}.
Under mild conditions, it can be shown that the invariant distribution of the continuous-time process is $\pi(\bbX) \propto p(\bbX)^{1/\tau}$~\cite{pavliotis_book}.
In particular, if $ \tau = 1$, then $\pi(\bbX) \propto p(\bbX)$.

The Euler-Maruyama discretization of~\eqref{eq:diffusion} gives rise to the unadjusted Langevin algorithm (ULA), which is an MCMC algorithm \cite{Roberts1996ExponentialCO}, described by the following discrete equation
\begin{equation}\label{eq:langevin}
	\bbX_{k+1} = \bbX_k + \epsilon \nabla_{\bbX_k}\log p(\bbX_k) + \sqrt{2\epsilon \tau}\, \bbZ_k,
\end{equation}
where $\bbZ_k\sim\ccalN(\bb0, \bbI)$.
In essence, ULA generates samples from a target distribution $p(\bbX)$ by iteratively moving in the direction of the gradient of the logarithm of the target density, known as the score function, and introducing noise to avoid local maxima.
Under some regularity conditions, the distribution of $\bbX_k$ converges to $p(\bbX)$ when $\epsilon \rightarrow 0$ and $k \rightarrow \infty$. 
Although this result is asymptotic, also non-asymptotic convergence results have been obtained under some conditions on the target distribution~\cite{dalalyan2019}.

% In practice, neither $\epsilon \rightarrow 0$ nor $T \rightarrow \infty$, so a Metropolis-Hastings acceptance/rejection step is used to ensure convergence, leading to the so-called Metropolis-adjusted Langevin algorithm (MALA)~\cite{Roberts1996ExponentialCO}.
% An alternative, proposed in~\cite{wellinglang}, is a time-inhomogeneous variant of~\eqref{eq:langevin}, i.e., defining a variable step size $\epsilon_t$.
% Thus, when $\epsilon_t$ decreases to zero for large $t$, the error becomes negligible and the acceptance/rejection step can be omitted.
% Although this result is asymptotic, also non-asymptotic convergence results have been obtained under some conditions on the target distribution~\cite{dalalyan2019}.

\vspace{0.2cm}
\noindent \textbf{Score-based generative modeling.} 
% \red{this part is not super clear, but I do not have a quick fix now. We can continue working on it while Ananthram takes his pass} 
It should be noted that the \textit{only} requirement for sampling from $p(\bbX)$ using this procedure is knowing the score function, which is unknown in general.
Hence, we consider denoising score matching~\cite{vincent2011connection} to estimate the score by training a neural network, known as a score network, that parameterizes the score.
In a nutshell, the method is as follows: given a training dataset $\{\bbX_i\}_{i=1}^n$ drawn from the distribution $p(\bbX)$, we first perturb the data at different scales with Gaussian kernels of variance $\{\sigma_l\}_{l=1}^L$ associated with each scale $l$.
This perturbation defines a distribution $q_{\sigma_l}(\tilde{\bbX})$ where $\tilde{\bbX}$ is the perturbed data.
Finally, the authors in~\cite{ermon2019} propose to estimate a joint score network $\bbs_{\theta}(\bbX, \sigma)$ via score matching, i.e., by minimizing the following loss
\begin{equation}\label{eq:loss}
\hspace{-0.4cm}\ccalL(\bbtheta) \!\!=\!\! \frac{1}{2L}\sum_{l=1}^L \lambda(\sigma_l)\mathbb{E}_{p(\bbX)q_{\sigma_l}(\tilde{\bbX}|\bbX)}\bigg[\bigg|\bigg|\bbs_{\theta}(\tilde{\bbX}, \sigma_l) + \frac{\tilde{\bbX} - \bbX}{\sigma_l^2}\bigg|\bigg|^2_2\bigg].
\end{equation}
where $\lambda(\sigma_l)$ is a pre-defined weight depending on $\sigma_l$.
After training, we can replace $\nabla_{\bbX_k}\log p(\bbX_k)$ in~\eqref{eq:langevin} by $\bbs_{\theta}(\bbX, \sigma)$ and generate samples from the target distribution $p(\bbX)$.
In Section~\ref{subsec:posterior}, we leverage this and use a pre-trained score network for the channel distribution.

%%%%%%%%%%%%%%------------
\vspace{-0.1in}
\subsection{Joint diffusion posterior sampling}
\label{subsec:posterior}

% Recall that our goal is to solve Problem~\ref{P:main} by sampling (approximately) from the joint posterior defined in~\eqref{eq:map} using ULA~\eqref{eq:langevin}.
% Therefore, we need to adapt the framework introduced in Section~\ref{subsec:langevindyn} as it does not apply directly to Problem~\ref{P:main}, mainly for three reasons.
% First, \emph{we do not seek to sample from $p(\bbX)$, but rather from the joint posterior $p(\bbX_D, \bbH|\bbY, \bbX_P)$.}
% Hence, we need to compute the score of the joint posterior distribution, i.e., the gradients with respect to $\bbX_D$ and $\bbH$.
% Both scores can be written after applying Bayes' rule as 

Recall our goal is to solve Problem~\ref{P:main} via ULA~\eqref{eq:langevin} by sampling from the joint posterior~\eqref{eq:map}. 
Hence, we must adapt the framework from Section~\ref{subsec:langevindyn} for several reasons. 
First, \emph{we are interested in sampling from the joint posterior $p(\bbX_D, \bbH|\bbY, \bbX_P)$, not just $p(\bbX)$.}
This necessitates computing the score of the joint posterior, specifically gradients with respect to $\bbX_D$ and $\bbH$, expressed as:
\begin{align}\label{E:score_function}
\hspace{-0.1cm}\nabla_{\bbX_D}\log p(\bbX_D, \bbH|\bbY, \bbX_P)  &= \nabla_{\bbX_D}\log p(\bbY|\bbH, \bbX) + \\\nonumber 
&\hspace{2.4cm}\nabla_{\bbX_D}\log p(\bbX_D), \\\nonumber
\hspace{-0.1cm}\nabla_{\bbH}\log p(\bbX_D, \bbH|\bbY, \bbX_P) &= \nabla_{\bbH}\log p(\bbY|\bbH, \bbX) + \\\nonumber
&\hspace{2.4cm}\nabla_{\bbH}\log p(\bbH)
\end{align}
where $\nabla_{\bbX_D}\log p(\bbY, \bbX_P|\bbH, \bbX_D)$ and $\nabla_{\bbX_D}\log p(\bbX_D)$ are the score functions of the likelihood and prior with respect to $\bbX_D$ respectively, while $\nabla_{\bbH}\log p(\bbY, \bbX_P|\bbH, \bbX_D)$ and $\nabla_{\bbH}\log p(\bbH)$ are the scores with respect to $\bbH$.
% Second, \emph{the gradient with respect to the prior of $\bbX_D$ is not well defined} given that $\bbX_D$ is a discrete variable.
% To circumvent this issue, we propose to leverage an \emph{annealing process,} which renders a continuous approximation for $\bbX_D$.
% Specifically, we define a sequence of noise levels\footnote{We use the $\bbX$ in the subscript to differentiate the annealing noise for $\bbX_D$ from the one used for $\bbH$.} $\{\sigma_{l, \bbX}\}_{l=1}^{L+1}$ such that $\sigma_{1, \bbX} > \sigma_{2, \bbX} > \cdots > \sigma_{L, \bbX} > 0$ and $\sigma_{L, \bbX} \approx 0$. 
% Then, at each level, we define a perturbed version of the true symbols $\bbX_D$
Second, \emph{computing the gradient for the prior of $\bbX_D$ is challenging due to its discrete nature}. 
To address this, we propose an \textit{annealing process}, creating a continuous approximation of $\bbX_D$.
We define a sequence of noise levels $\{\sigma_{l, \bbX}\}_{l=1}^{L}$, with decreasing values so that $\sigma_{L, \bbX} \approx 0$.
Then, at each level, we introduce a perturbed version of the true symbols $\bbX_D$
\begin{equation}\label{eq:pert_symbs}
    \tilde{\bbX}_{D,l} = \bbX_D + \bbN_{\bbX, l},
\end{equation}
where $\bbN_{\bbX, l} \sim \mathcal{CN}(0, \sigma_{l, \bbX}^2\bbI)$. 
Hence, we utilize the perturbed symbol set $\tilde{\bbX}_D$ instead of the true variable $\bbX_D$, enabling the definition of a continuous score prior. 
Therefore, as noise levels decrease, $\tilde{\bbX}_D$ progressively converges to $\bbX_D$. 
Although this is similar to the concept introduced in Section~\ref{subsec:langevindyn}, the ultimate goal is different, as we aim to create a continuous approximation of the discrete variable $\tilde{\bbX}_D$.
Furthermore, we \emph{lack a closed-form expression for the score of $\bbH$, prompting the use of a score network parameterization and training via denoising score matching}, as presented in Section~\ref{subsec:langevindyn}.

In a nutshell, the algorithm follows these steps: initialization of $\tilde{\bbX}_{D,0}$ and $\tilde{\bbH}_0$ randomly. 
After that, it follows the score function of the joint posterior density of perturbed variables, starting with high $\sigma_{1, \bbX}$ and $\sigma_{1, \bbH}$, progressively reducing to $\sigma_{L, \bbX} \approx \sigma_{L, \bbH} \approx 0$. 
At early noise levels, the likelihood term directs the dynamics toward an estimate mainly driven by the measurements, while in later noise levels, the prior refines the estimate, as explained further in~\cite{zilberstein2022annealed}. 
Annealing benefits are threefold: it is used to train the score network via score-matching, it enhances dynamic mixing, and it allows for discrete-to-continuous variable approximation. 
Consequently, we apply annealing to both $\bbH$ and $\bbX_D$ to harness the first two benefits for the former and the last two for the latter.
% \red{We talk about using annealing for the discrete nature of X, but H is not discrete and we still use annealing there. I know that annealing can be used to improve the convergence to the stationary distribution, but do we mention any of this anywhere? Maybe mention it here something like "Consequently, we apply annealing to both ...." Otherwise, the reader might wonder why are we annealing an already continuous variable.}
With this high-level understanding, we now detail the terms in~\eqref{E:score_function} and provide a step-by-step algorithm description.

\vspace{2mm}
\noindent \textbf{Score functions.} The score functions are computed with respect to the perturbed variables.
Thus, our model for running the dynamic in~\eqref{eq:langevin} is described by the following joint posterior distribution
\begin{equation}\label{eq:forwardmodel_noise}
    p(\tilde{\bbX}_{D,l}, \tilde{\bbH}_l| \bbY, \bbX_P) \propto p( \bbY, \bbX_P |\tilde{\bbX}_{D,l}, \tilde{\bbH}_l)p(\tilde{\bbX}_{D,l})p(\tilde{\bbH}_{l}).
\end{equation}
Hence,  we need to compute four score functions~\eqref{E:score_function}, two corresponding to the score of the likelihood terms, and two for the score prior.

\vspace{1mm}
\noindent \emph{i) Score of the likelihood term of $\tilde{\bbX}_{D,l}$}: Under the new model in~\eqref{eq:forwardmodel_noise}, the score of the likelihood term for $\tilde{\bbX}_{D,l}$ is given by $p(\bbY|\tilde{\bbX}_l, \tilde{\bbH}_l, \bbX_P) = p(\bbZ - \tilde{\bbH}_l {\bbN}_{\bbX,l}|\tilde{\bbX}_{D,l})$, which is not Gaussian: although $p({\bbN}_{\bbX,l})$ is a Gaussian distribution, after conditioning on the perturbed variable the conditional distribution $p({\bbN}_{\bbX,l}|\tilde{\bbX}_{D,l})$ is no longer Gaussian. 
To circumvent this, we consider the following \textit{approximation}.
First, notice that when $\sigma_{l, \bbX}$ is small, then $\nabla_{\tilde{\bbX}_D}\log p(\bbY|\tilde{\bbX}_{D,l}, \tilde{\bbH}_l, \bbX_P) \approx \nabla_{\bbX_D}\log p(\bbY_D|\bbX_D, \tilde{\bbH}_l) = \frac{\tilde{\bbH}^{\text{H}}_l(\bbY_D - \tilde{\bbH}_l\bbX_D)}{\sigma_0^2}$, with $\tilde{\bbH}^{\text{H}}_l$ denoting the conjugate transpose of $\bbH$; this is only correct when $\sigma_{l, \bbX} \rightarrow 0$ (recall that for $\sigma_{l, \bbX} = 0$, the gradient is not defined).
Second, given that the approximation is not valid for high $\sigma_{l, \bbX}$, we add a correction term as follows
%
% \begin{equation}\label{eq:guidance_x}
%     \nabla_{\tilde{\bbX}_D}\log p(\bbY|\tilde{\bbX}_l, \tilde{\bbH}_l, \bbX_P) \approx \frac{\tilde{\bbH}_l^{\text{H}}(\bbY_D - \tilde{\bbH}_l\tilde{\bbX}_D)}{\sigma_0^2 + \sigma_{l, \bbX}^2 \tilde{\bbH}_l^{\text{H}}\tilde{\bbH}_l}.
% \end{equation}
\begin{align}\label{eq:guidance_x}
    \nabla_{\tilde{\bbX}_D}\log p(\bbY|\tilde{\bbX}_{D,l}, \tilde{\bbH}_l, \bbX_P) &\approx \\\nonumber
    &\hspace{-1cm}\tilde{\bbH}_l^{\text{H}}(\sigma_0^2\bbI + \sigma_{l, \bbX}^2 \tilde{\bbH}_l^{\text{H}}\tilde{\bbH}_l)^{-1}(\bbY_D - \tilde{\bbH}_l\tilde{\bbX}_{D,l}).
\end{align}
Intuitively, the approximation in~\eqref{eq:guidance_x} assumes that the annealing and measurement noise are independent.
Although this approximation entails a worse symbol error rate (SER) than the one introduced in~\cite{zilberstein2022annealed},
% \red{dide also use an approximation in~\cite{zilberstein2022annealed} or was it an exact computation? ... this sentence reads as it was an approximation}
it provides a good approximation that allows one to reduce the computational burden of the algorithm.

\vspace{1mm}
\noindent \emph{ii) Score of the likelihood term of $\tilde{\bbH}_l$:} Similar to the approximation in~\eqref{eq:guidance_x}, we approximate the score of likelihood of $\tilde{\bbH}_l$ as follows
\begin{equation}\label{eq:guidance_H}
    \hspace{-0.3cm}\nabla_{\tilde{\bbH}_l}\log p(\bbY|\tilde{\bbX}_{D,l}, \tilde{\bbH}_l, \bbX_P) \approx \frac{(\bbY - \tilde{\bbH}_l[\tilde{\bbX}_{D,l}, \bbX_P])[\tilde{\bbX}_{D,l}, \bbX_P]^{\text{H}}}{\sigma_0^2 + \sigma_{l, \bbH}^2}
\end{equation}
Notice that this term considers both the pilots and the data in contrast to the score of the likelihood of $\tilde{\bbX}_D$ in~\eqref{eq:guidance_x}

\vspace{1mm}
\noindent \emph{iii) Score of the annealed prior of $\tilde{\bbX}_{D,l}$:} The score function can be related to the MMSE denoiser through Tweedie's identity~\cite{TweedieIdent} as follows
\begin{equation}\label{eq:prior}
	\nabla_{\tilde{\bbX}_{D, l}}\log p(\tilde{\bbX}_{D, l}) =  \frac{\mathbb{E}_{\sigma_{l, \bbX}}[\bbX_D|\tilde{\bbX}_{D, l}] - \tilde{\bbX}_{D, l}}{\sigma_{l, \bbX}^2}.
\end{equation}
In particular, the conditional expectation can be calculated elementwise as
{\footnotesize\begin{align}\label{E:mixed_gaussian}
    \hspace{-0.3cm}\mathbb{E}_{\sigma_{l, \bbX}}[x_j|[\tilde{\bbX}_{D, l}]_{j,p}] &= \frac{1}{Z}\sum_{x_k \in \ccalX} x_k \exp\bigg(\frac{-||[\tilde{\bbX}_{D, l}]_{j,p} - x_k||^2}{2\sigma_{l, \bbX}^2}\bigg),
\end{align}}
where $Z = \sum_{x_k \in \ccalX} \exp\Big(\frac{-||[\tilde{\bbX}_{D, l}]_{j,p} - x_k||^2}{2\sigma_{l, \bbX}^2}\Big)$ and $j=1,\cdots, N_u$. 

\vspace{1mm}
\noindent \emph{iv) Score of the annealed prior of $\tilde{\bbH}_{l}$:} Finally, the score prior for $\tilde{\bbH}_l$ is parameterized by a score network, and trained using denoising score matching, which was introduced in Section~\ref{subsec:langevindyn}. 
Hence, 
\begin{align}\label{E:score_network}
    \nabla_{\tilde{\bbH}_{l}}\log p(\tilde{\bbH}_{l}) = \bbs_{\boldsymbol{\theta}}(\tilde{\bbH}_l, \sigma_{l, \bbH}),
\end{align}
where the network $\bbs_{\boldsymbol{\theta}}(.)$ is trained by minizing the loss function in~\eqref{eq:loss}.
\begin{algorithm}[t]
    \footnotesize
    \caption{Annealed Langevin for JED-MAP}\label{alg}
	\begin{algorithmic}
		\Require $T, L, \{\sigma_{l, \bbX}, \sigma_{l, \bbH}\}_{l=1}^L, \epsilon_\bbX, \epsilon_\bbH, \sigma_0, \bbY, \bbX_P , \tau_\bbX, \tau_\bbH$
		\State Initialize $\tilde{\bbX}_{D_{t=0,l=1}}, \tilde{\bbH}_{t=0,l=1}$ randomly
		\For{$l = 1\; \text{to}\;  L$}
              \State Compute $\epsilon_{l, \bbX} = \epsilon_\bbX \cdot \Big(\frac{\sigma_{l, \bbX}}{\sigma_{L, \bbX}}\Big)^2, \epsilon_{l, \bbH} =  \epsilon_\bbH \cdot \Big(\frac{\sigma_{l, \bbH}}{\sigma_{L, \bbH}}\Big)^2$
		\For{$k = 0\; \text{to}\; T-1$}
			\State Draw $\bbZ_{k} \sim \ccalN(0, \bbI)$

            \State Compute $\nabla_{\tilde{\bbX}_{D_{k,l}}}\log p(\bbY|\tilde{\bbX}_{D_{k,l}}, \tilde{\bbH}_{k, l}, \bbX_P)$ as in~\eqref{eq:guidance_x}

            \State Compute $\nabla_{\tilde{\bbX}_{D_{k,l}}}\log p(\tilde{\bbX}_{D_{k,l}})$
            as in~\eqref{eq:prior}
            %\gets  \frac{ \mathbb{E}[\bbx|\tilde{\bbx}_{k,l}] - \tilde{\bbx}_{k,l}}{\sigma_l^2}$
            
            \State Compute $\nabla_{\tilde{\bbX}_{D_{k,l}}}\log p(\tilde{\bbX}_{D_{k,l}}, \tilde{\bbH}_{k, l}|\bbY, \bbX_P)$
            as in~\eqref{E:score_function}
   
            \State $\tilde{\bbX}_{D_{k+1, l}} = \tilde{\bbX}_{D_{k, l}}  + \epsilon_{l, \bbX} \nabla_{\bbX_{D_{k, l}}}\hspace{-0.1cm}\log p(\tilde{\bbX}_{D_{k, l}}, \tilde{\bbH}_{k, l}|\bbY, \bbX_P) $
            \State ~~~~$ + \sqrt{2\epsilon_{l, \bbX} \; \tau_{\bbX}}\, \bbZ_{k}$    
            
            \State Compute $\nabla_{\tilde{\bbH}_{k, l}}\log p(\bbY|\tilde{\bbX}_{D_{k+1,l}}, \tilde{\bbH}_{k, l}, \bbX_P)$ as in~\eqref{eq:guidance_H}

            \State Compute $\nabla_{\tilde{\bbH}_{l}}\log p(\tilde{\bbH}_{l})$ as in~\eqref{E:score_network}
            
            \State Compute $\nabla_{\tilde{\bbH}_{k, l}}\log p(\tilde{\bbX}_{D_{k+1,l}}, \tilde{\bbH}_{k,l}|\bbY, \bbX_P)$
            as in~\eqref{E:score_function}

            \State $\tilde{\bbH}_{k+1, l} = \tilde{\bbH}_{k, l} + \epsilon_{l, \bbH} \nabla_{\tilde{\bbH}_{k, l}}\hspace{-0.15cm}\log p(\tilde{\bbX}_{D_{k+1,l}}, \tilde{\bbH}_{k, l}|\bbY, \bbX_P) $
            \State ~~~$+ \sqrt{2\epsilon_{l, \bbH} \; \tau_{\bbH}}\, \bbZ_{k}$      
            % \red{weird subscript in the epsilon, is it correct?}
		\EndFor
		\EndFor \\
	\Return $\hat{\bbX}_D = \tilde{\bbX}_{D_{T,L}}, \hat{\bbH} = \tilde{\bbH}_{T, L}$
	\end{algorithmic}
\end{algorithm}
% \vspace{-0.1in}

%%%%%%%%%%   F   I   G   U   R    E
\begin{figure*}[t]
	\begin{subfigure}{.33\textwidth}
    	\centering
    	\includegraphics[width=1\textwidth]{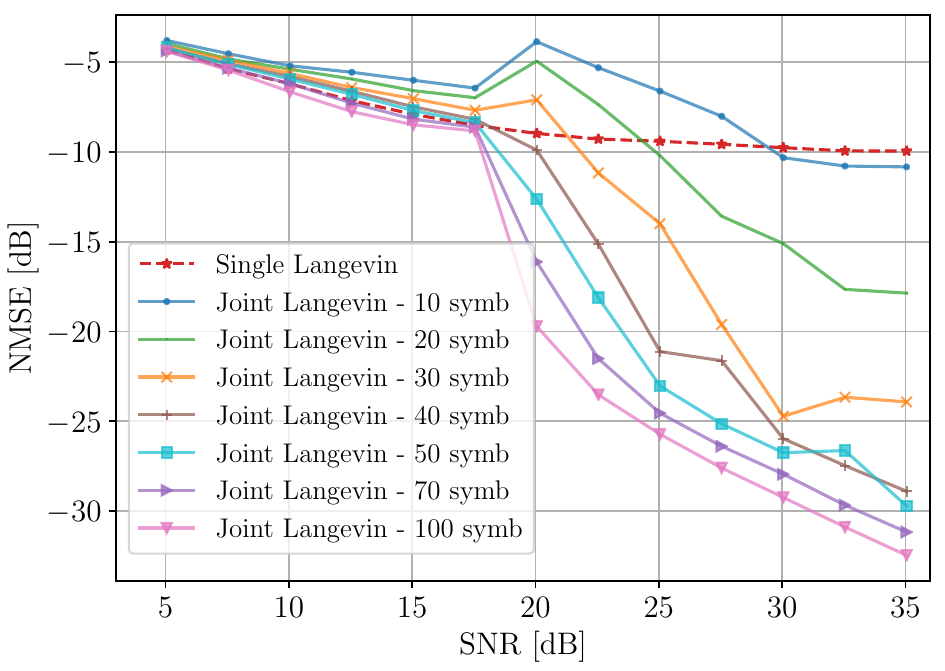}
    	\vspace{-0.2in}
    	\caption{}
    	\label{fig:nsymbs-comparison}
	\end{subfigure}
	\begin{subfigure}{.33\textwidth}
    	\centering
    	\includegraphics[width=1\textwidth]{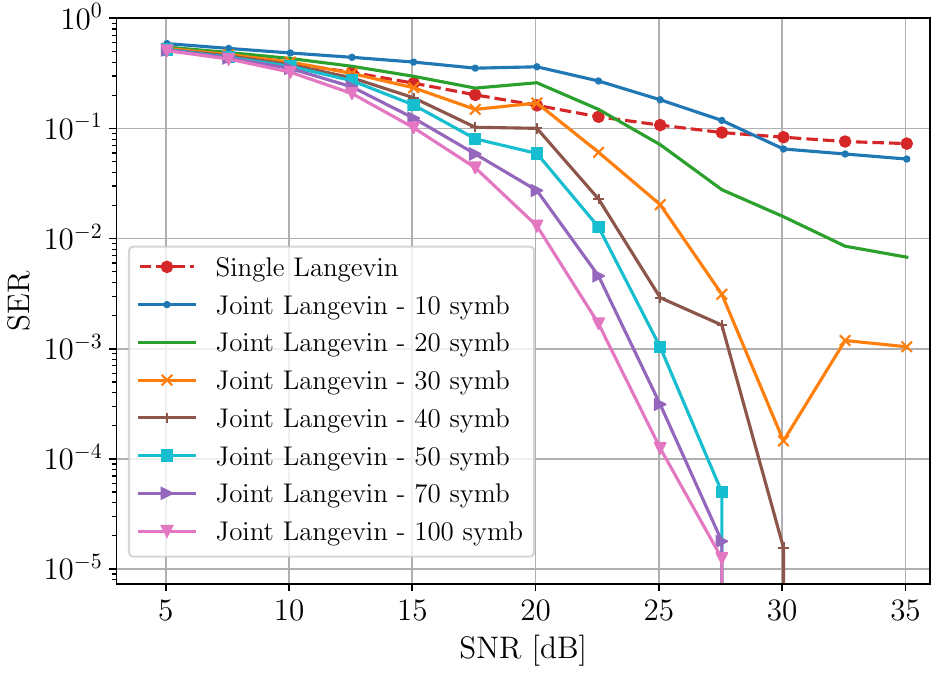}
    	\vspace{-0.2in}
    	\caption{}
    	\label{fig:nser-comparison}
	\end{subfigure}
	\begin{subfigure}{.33\textwidth}
    	\centering
    	\includegraphics[width=1\textwidth]{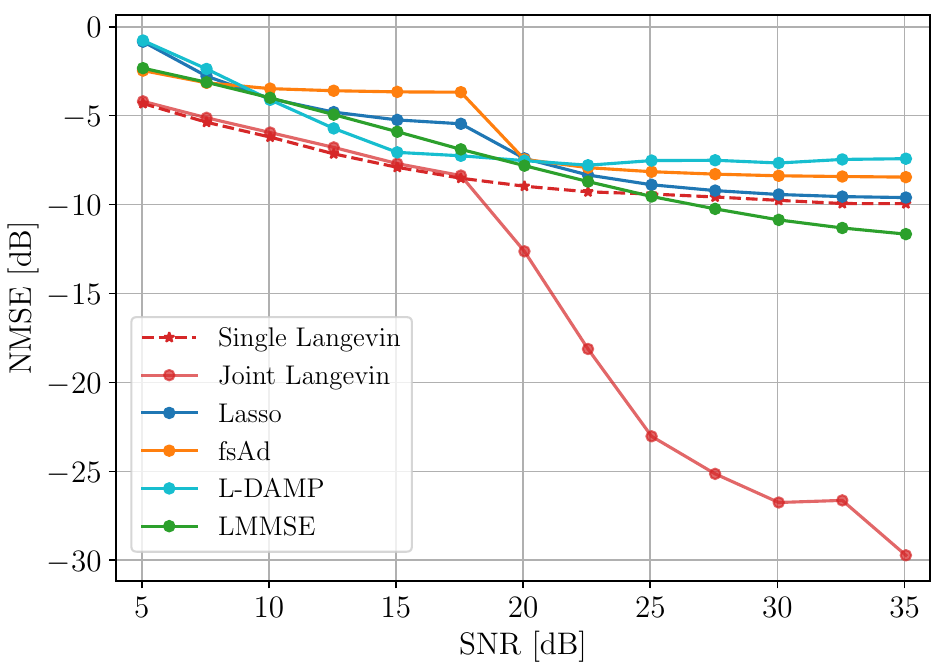}
    	\vspace{-0.2in}
    	\caption{}
    	\label{fig:baseline-comparison}
	\end{subfigure}%
	\vspace{-0.08in}
	\caption{ {\footnotesize Performance analysis of our proposed method as a function of SNR for a 3GPP channel model.
	(a), (b)~NMSE and SER respectively for a fixed number of pilot symbols $P = 30$ and different number of data symbols, $D = \{10, 20, 30, 40, 50, 70, 100\}$. 
	(c)~Comparison with different baseline methods, for $P=30$ and $D=50$.}}
	\vspace{-0.1in}
	\label{fig_results}
\end{figure*}
%%%%%%%%%%%%%%%%%%%%%

\vspace{2mm}
\noindent {\bf Algorithm.} 
The algorithm to generate estimates $\hat{\bbH}$ and $\hat{\bbX}_D$ by sampling from the (approximate) joint posterior $p(\bbX_D, \bbH|\bbY,\bbX_P)$ is shown in Algorithm~\ref{alg}.
From the update steps, we observe that the main advantage of performing joint sampling is to reuse data symbols as pilots, which entails two benefits.
First, it allows sending more data with the same data rate and maintaining a given system performance.
Second, it reduces the channel estimation error as long as the estimation of the symbols is good.
Hence, one might expect to see an improvement at higher SNRs.
Finally, from an implementation perspective, the selection of the hyperparameters -- the number of levels of noise as well as the step sizes -- is key for the stability of the algorithm.

\vspace{1mm}
\noindent {\bf Computational complexity.}  
The complexity of Algorithm~\ref{alg} is the combination of the complexities of each dynamic.
The dynamic with respect to the variable $\bbX_D$ yields a complexity where the main bottleneck is the matrix inversion in~\eqref{eq:guidance_x}.
Hence, the complexity of computing the guidance term and the score prior is $\ccalO(N_u^3 + KN_u)$.
% \red{Out of curiosity ... if this is still cubic, how is this better than the SVD that we are trying to avoid? Nothing to change in the paper, but let's discuss it a bit over slack.}
A deeper analysis of the complexity of this method can be found in~\cite{zilberstein2022annealed}.
For the case of $\bbH$, the heavier computation is carried by the score network.
An analysis of the computation complexity of this method can be found in~\cite{arvinte2022}.

%%%%%%%%%%%%%%%%%%%%%%%%%%%%%%%%%%%%%
%%%%%%%%Results%%%%%%%%%%%%%%%%%%%%%%
%%%%%%%%%%%%%%%%%%%%%%%%%%%%%%%%%%%%%
\vspace{-0.1in}
\section{Numerical Experiments}
\label{S:numerical_experiments}

\vspace{-0.1in}
% In this section, we present the results of our proposed method.\footnote{Code to replicate the numerical experiments can be found at \url{https://github.com/nzilberstein/Langevin-joint-channel}}
% We start by presenting the channel model and the simulation setup.
% In the first experiment, we analyze the channel estimation error of the proposed method as a function of the signal-to-noise ratio (SNR) for different number of symbols $\bbX_D$ and a fixed number of pilots, and compare it to the case when no symbols are reused as pilots.
% In the second experiment, we fixed the number of symbols and consider different number of pilots.
% Finally, we compare our proposed method with other baseline methods in the task of channel estimation.
In this section, we evaluate the performance of our method.
%This section presents our method's results.
\footnote{Code for reproducing experiments is available at \url{https://github.com/nzilberstein/Langevin-joint-channel}} 
We begin by introducing the channel model and simulation setup. 
The first experiment assesses channel estimation error versus signal-to-noise ratio (SNR), varying the number of symbols $\bbX_D$ with a fixed pilot count, comparing against the case with no symbol reuse as pilots. 
In the second experiment, we consider the same setting as experiment one and evaluate the SER as a function of SNR.
Finally, we benchmark our approach against baseline methods in channel estimation.

\vspace{1mm}
\noindent {\bf Channel model and simulation settings.} We consider a channel model that is representative of the 3GPP 3D MIMO channel model~\cite{3gpp}, 
as implemented in the QuaDRiGa channel simulator~\cite{quadriga}.
We consider a base station with an $8\times8$ half-wavelength spacing ($N_r=64$), single-polarization antenna array at a height of \SI{20}{\meter}.
We assume that the BS covers a sector of radius \SI{500}{\meter} and $N_u=32$ single-polarization omni-directional antennas users are dropped randomly in the coverage area.
Moreover, users are NLOS and indoors.
The carrier frequency is \SI{3.5}{\giga\hertz}, and  each subcarrier has a \SI{100}{\mega\hertz} bandwidth. The spacing within subcarriers is \SI{30}{\kilo\hertz}.
For the score network, we use the parameterization proposed in~\cite{ermon2020}, and minimize the loss in~\eqref{eq:loss}.
We consider a training dataset with $7000$ channels, and a validation set of $2500$ channels.
We evaluate the performance in a batch size of $50$ matrices.
For the pilots, we generate $P$ QPSK symbols.
The parameters of the Algorithm~\ref{alg} are $L = 2311, T = 3, \epsilon_{\bbH} = 1 \times 10^{-10}, \tau_{\bbH} = 1\times 10^{-3}$ and $[\sigma_{1,\bbH}, \sigma_{L, \bbH}] = [30, 0.001]$; lastly, for the dynamic with respect to $\bbX_D$, we use two different values depending on the SNR: for low SNR (\SI{5}{\decibel}-\SI{15}{\decibel}) we consider $\epsilon_{\bbX} = 1 \times 10^{-4}, \tau_{\bbX} = 0.5$ and $[\sigma_{1,\bbX}, \sigma_{L, \bbX}] = [0.6,0.01 ]$, while for all other SNR values we fixed $\epsilon_{\bbX} = 4 \times 10^{-5}, \tau_{\bbX} = 0.1$ and $[\sigma_{1,\bbX}, \sigma_{L, \bbX}] = [0.8,0.01]$. 
Lastly, the noise variance $\sigma_0^2$ changes for different levels of SNR.
If computationally limiting, $L$ can be reduced by considering higher-order Langevin dynamics; see~\cite{zilberstein2023solving} for an analysis of the case of channel sampling.

\vspace{1mm}
\noindent{\bf NMSE performance for varying number of symbols.}
In this first experiment, we analyze the performance of our proposed algorithm for channel estimation with different values for $D$, i.e., symbols to reuse as pilots, and fixing the value of $P = 30$.
The comparison is shown in Fig.~\ref{fig:nsymbs-comparison}, where we named \textit{single Langevin} the channel sampling for $D = 0$, proposed in~\cite{arvinte2022}.
As expected, performance improves with an increasing number of data symbols for pilot reuse. 
Notably, when $D > 30$, our method excels for SNRs above \SI{20}{\decibel}, and for $D > 50$, it outperforms single Langevin across all SNRs. 
Overall, superior performance is evident for SNRs higher than \SI{20}{\decibel}.
% As it is expected, we observe that the performance improves when considering more data symbols to reuse.
% In particular, when $D > 30$, we observe a clear improvement in performance for SNRs higher than \SI{20}{\decibel}, while for $D > 50$, our proposed method outperforms the single Langevin for all SNRs.
% Lastly, from the plot we notice that the performance is superior for SNRs bigger than \SI{17.5}{\decibel}.

\vspace{1mm}
\noindent{\bf SER performance for varying number of symbols.}
In the same experimental conditions as before (where $P = 30$), we examine the SER as a function of SNR, with varying values of $D$. The results are presented in Fig.~\ref{fig:nser-comparison}. Similar to the behavior observed in channel estimation error, we notice an improvement in SER as $D$ increases. Notably, when $D > 30$, our proposed algorithm outperforms the single Langevin method across all SNRs. These findings suggest that the limitation in NMSE performance in our algorithm stems from symbol sampling performance. In high SNR scenarios, accurate estimation of $\bbX_D$ leads to precise joint variable estimation. However, in lower SNR situations, there is not a significant advantage in considering symbol reuse.
% We now show the NMSE for a fixed value of $D$ and different values of $P$, i.e., different number of pilots. 
% We consider $D = 50$.
% The results are illustrated in Fig.~\ref{fig:npilots-comparison}.
% First, we observe that our proposed method outperforms the single Langevin for all SNRs when the number of pilots is 24, 26 and 30, which are the most interesting cases as the pilot overhead is reduced with respect to the total number of users. 
% Then, when there are more pilots available (37 and 64), our method outperforms the single Langevin only for SNRs bigger than \SI{20}{\decibel}.
% Based on the last two experiments, we can conclude that the limitation in performance of our proposed algorithm is due to the performance of the sampling of the symbols.
% Hence, when the estimation of $\bbX_D$ is accurate, then the joint sampling will generate an accurate estimation of both variables, which is the case for SNRs higher than \SI{18}{\decibel}.
% This conclusion matches with the performance of the detector presented in~\cite{zilberstein2022annealed}.

\vspace{1mm}
\noindent{\bf Performance comparison with baseline methods.}
In our third experiment, we compare our proposed algorithm to several baseline methods: LASSO~\cite{lasso}, fsAD~\cite{fsad}, L-MMSE~\cite{nayebi2017semi}, L-DAMP~\cite{ldamp1}, and Langevin using only pilots~\cite{arvinte2022}. 
We set $P = 30$ for all baseline algorithms and $D = 50$ for our method. 
Results in Fig.~\ref{fig:baseline-comparison} reveal that our approach consistently outperforms all baselines across all SNRs. Particularly noteworthy is its superior performance, exceeding other methods by several orders of magnitude for SNRs above \SI{15}{\decibel}.
% In this third experiment, we compare our proposed algorithms with the following baseline methods: LASSO~\cite{lasso}, fsAD~\cite{fsad}, L-MMSE~\cite{nayebi2017semi} and Langevin using only pilots~\cite{arvinte2022}.
% We consider $P = 30$ for all the baseline algorithms, and $D = 50$ for our proposed method.
% The results are shown in Fig.~\ref{fig:baseline-comparison}, where we plot the NMSE as a function of SNR.
% First, we observe that our proposed method outperforms all the other baseline for all SNR.
% And in particular, for SNRs higher than \SI{15}{\decibel}, it does it by several order of magnitudes.

\vspace{-0.1in}
\section{Conclusions}
\vspace{-0.1in}
We introduced an algorithm for joint massive MIMO channel estimation and data detection using diffusion models. 
Our approach leverages the reverse processes to generate samples for solving the MAP-JED optimization problem, accommodating both discrete and continuous variable priors. 
Simulations reveal our method's superior performance over baselines while managing pilot overhead. 
Future work aims to explore streamlined diffusion models, reducing hyperparameters and accelerating sampling, and enhancing performance in lower SNR scenarios.

%%%%%%%%%%%%%%%%%%%%%%%%%%%%%%%%%%%%%%%%%%%%%%%%%%%%%%%%%%%%

\bibliographystyle{IEEEbib}
\bibliography{citations}

\begin{thebibliography}{10}

\bibitem{mimoreview1}
Shaoshi Yang and Lajos Hanzo,
\newblock ``Fifty years of {MIMO} detection: The road to large-scale {MIMOs},''
\newblock {\em IEEE Commun. Surveys Tut.}, vol. 17, no. 4, pp. 1941--1988,
  2015.

\bibitem{6g}
Khaled~B. Letaief, Wei Chen, Yuanming Shi, Jun Zhang, and Ying-Jun~Angela
  Zhang,
\newblock ``The roadmap to {6G}: {AI} empowered wireless networks,''
\newblock {\em IEEE Commun. Mag.}, vol. 57, no. 8, pp. 84--90, 2019.

\bibitem{Pia2017MixedintegerQP}
Alberto~Del Pia, Santanu~S. Dey, and M.~Molinaro,
\newblock ``Mixed-integer quadratic programming is in {NP},''
\newblock {\em Mathematical Programming}, vol. 162, pp. 225--240, 2017.

\bibitem{balevi2020high}
Eren Balevi, Akash Doshi, Ajil Jalal, Alexandros Dimakis, and Jeffrey~G
  Andrews,
\newblock ``High dimensional channel estimation using deep generative
  networks,''
\newblock {\em IEEE J. Sel. Areas Commun.}, vol. 39, no. 1, pp. 18--30, 2020.

\bibitem{vikalo2006efficient}
Haris Vikalo, Babak Hassibi, and Petre Stoica,
\newblock ``Efficient joint maximum-likelihood channel estimation and signal
  detection,''
\newblock {\em IEEE Trans. Wireless Commun.}, vol. 5, no. 7, pp. 1838--1845,
  2006.

\bibitem{yi2020deep}
Xuemei Yi and Caijun Zhong,
\newblock ``Deep learning for joint channel estimation and signal detection in
  {OFDM} systems,''
\newblock {\em IEEE Commun. Lett.}, vol. 24, no. 12, pp. 2780--2784, 2020.

\bibitem{castaneda2017vlsi}
Oscar Castaneda, Tom Goldstein, and Christoph Studer,
\newblock ``{VLSI} designs for joint channel estimation and data detection in
  large {SIMO} wireless systems,''
\newblock {\em IEEE Trans. Circuits Syst. I Regul. Pap.}, vol. 65, no. 3, pp.
  1120--1132, 2017.

\bibitem{song2021joint}
Haochuan Song, Tom Goldstein, Xiaohu You, Chuan Zhang, Olav Tirkkonen, and
  Christoph Studer,
\newblock ``Joint channel estimation and data detection in cell-free massive
  {MU-MIMO} systems,''
\newblock {\em IEEE Trans. Wireless Commun.}, vol. 21, no. 6, pp. 4068--4084,
  2021.

\bibitem{song2021soft}
Haochuan Song, Xiaohu You, Chuan Zhang, and Christoph Studer,
\newblock ``Soft-output joint channel estimation and data detection using deep
  unfolding,''
\newblock in {\em IEEE Inf. Theory. Workshop}. IEEE, 2021, pp. 1--5.

\bibitem{song2021scorebased}
Yang Song, Jascha Sohl-Dickstein, Diederik~P Kingma, Abhishek Kumar, Stefano
  Ermon, and Ben Poole,
\newblock ``Score-based generative modeling through stochastic differential
  equations,''
\newblock in {\em Intl. Conf. Learn. Repr. (ICLR)}, 2021.

\bibitem{kawar2022denoising}
Bahjat Kawar, Michael Elad, Stefano Ermon, and Jiaming Song,
\newblock ``Denoising diffusion restoration models,''
\newblock in {\em Advances in Neural Inf. Process. Syst. (NeurIPS)}, 2022.

\bibitem{chung2023diffusion}
Hyungjin Chung, Jeongsol Kim, Michael~Thompson Mccann, Marc~Louis Klasky, and
  Jong~Chul Ye,
\newblock ``Diffusion posterior sampling for general noisy inverse problems,''
\newblock in {\em Intl. Conf. Learn. Repr. (ICLR)}, 2023.

\bibitem{zilberstein2022annealed}
Nicolas Zilberstein, Chris Dick, Rahman Doost-Mohammady, Ashutosh Sabharwal,
  and Santiago Segarra,
\newblock ``Annealed {L}angevin dynamics for massive {MIMO} detection,''
\newblock {\em IEEE Trans. Wireless Commun.}, vol. 22, no. 6, 2023 (Online Nov
  2022).

\bibitem{zilberstein2023}
Nicolas Zilberstein, Chris Dick, Rahman Doost-Mohammady, Ashutosh Sabharwal,
  and Santiago Segarra,
\newblock ``Accelerated massive {MIMO} detector based on annealed underdamped
  {L}angevin dynamics,''
\newblock in {\em IEEE Intl. Conf. Acoust., Speech and Signal Process.
  (ICASSP)}, 2023.

\bibitem{arvinte2022}
Marius Arvinte and Jonathan~I. Tamir,
\newblock ``{MIMO} channel estimation using score-based generative models,''
\newblock {\em IEEE Trans. Wireless Commun.}, vol. 22, no. 6, 2023 (Online Nov
  2022).

\bibitem{levac2022accelerated}
Brett Levac, Ajil Jalal, and Jonathan~I Tamir,
\newblock ``Accelerated motion correction for {MRI} using score-based
  generative models,''
\newblock {\em arXiv preprint arXiv:2211.00199}, 2022.

\bibitem{Chung_2023_CVPR}
Hyungjin Chung, Jeongsol Kim, Sehui Kim, and Jong~Chul Ye,
\newblock ``Parallel diffusion models of operator and image for blind inverse
  problems,''
\newblock in {\em IEEE Conf. Comp. Vision Pattern Recognit. (CVPR)}, June 2023,
  pp. 6059--6069.

\bibitem{pavliotis_book}
Grigorios~A. Pavliotis,
\newblock {\em Stochastic Processes and Applications: Diffusion Processes, the
  Fokker-Planck and Langevin Equations},
\newblock Springer, 2014.

\bibitem{Roberts1996ExponentialCO}
Gareth~O. Roberts and Richard~L. Tweedie,
\newblock ``Exponential convergence of {L}angevin distributions and their
  discrete approximations,''
\newblock {\em Bernoulli}, vol. 2, pp. 341--363, 1996.

\bibitem{dalalyan2019}
Arnak~S. Dalalyan and Avetik Karagulyan,
\newblock ``User-friendly guarantees for the {L}angevin {M}onte {C}arlo with
  inaccurate gradient,''
\newblock {\em Stoch. Process. Their Appl.}, vol. 129, no. 12, pp. 5278--5311,
  2019.

\bibitem{vincent2011connection}
Pascal Vincent,
\newblock ``A connection between score matching and denoising autoencoders,''
\newblock {\em Neural Comput.}, vol. 23, no. 7, pp. 1661--1674, 2011.

\bibitem{ermon2019}
Yang Song and Stefano Ermon,
\newblock ``Generative modeling by estimating gradients of the data
  distribution,''
\newblock in {\em Advances in Neural Inf. Process. Syst. (NeurIPS)}, 2019, p.
  11918–11930.

\bibitem{TweedieIdent}
Bradley Efron,
\newblock ``Tweedie’s formula and selection bias,''
\newblock {\em Journal of the American Stat. Association}, vol. 106, no. 496,
  pp. 1602--1614, 2011.

\bibitem{3gpp}
3{GPP},
\newblock ``Study on 3-{D} channel model for {LTE},''
\newblock Tech. {R}ep. 36.873, 2015.

\bibitem{quadriga}
Stephan Jaeckel, Leszek Raschkowski, Kai Börner, and Lars Thiele,
\newblock ``{Q}ua{DR}i{G}a: A 3-{D} multi-cell channel model with time
  evolution for enabling virtual field trials,''
\newblock {\em IEEE Trans. Antennas Propag.}, vol. 62, no. 6, pp. 3242--3256,
  2014.

\bibitem{ermon2020}
Yang Song and Stefano Ermon,
\newblock ``Improved techniques for training score-based generative models,''
\newblock {\em arXiv preprint arXiv:2006.09011}, 2020.

\bibitem{zilberstein2023solving}
Nicolas Zilberstein, Ashutosh Sabharwal, and Santiago Segarra,
\newblock ``Solving linear inverse problems using higher-order annealed
  {L}angevin diffusion,''
\newblock {\em arXiv preprint arXiv:2305.05014}, 2023.

\bibitem{lasso}
Kiran Venugopal, Ahmed Alkhateeb, Nuria González~Prelcic, and Robert~W. Heath,
\newblock ``Channel estimation for hybrid architecture-based wideband
  millimeter wave systems,''
\newblock {\em IEEE J. Sel. Areas Commun.}, vol. 35, no. 9, pp. 1996--2009,
  2017.

\bibitem{fsad}
Badri~Narayan Bhaskar, Gongguo Tang, and Benjamin Recht,
\newblock ``Atomic norm denoising with applications to line spectral
  estimation,''
\newblock {\em IEEE Trans. Signal Process.}, vol. 61, no. 23, pp. 5987--5999,
  2013.

\bibitem{nayebi2017semi}
Elina Nayebi and Bhaskar~D Rao,
\newblock ``Semi-blind channel estimation for multiuser massive {MIMO}
  systems,''
\newblock {\em IEEE Trans. Signal Process.}, vol. 66, no. 2, pp. 540--553,
  2017.

\bibitem{ldamp1}
Chris Metzler, Ali Mousavi, and Richard Baraniuk,
\newblock ``Learned {D-AMP}: Principled neural network based compressive image
  recovery,''
\newblock {\em Advances in Neural Inf. Process. Syst. (NeurIPS)}, vol. 30,
  2017.

\end{thebibliography}

\end{document}